**Direct atomic layer deposition of ultrathin aluminium oxide on monolayer MoS₂ exfoliated on gold: the role of the substrate**

*Emanuela Schilirò, Raffaella Lo Nigro\*, Salvatore E. Panasci, Simonpietro Agnello, Marco Cannas, Franco M. Gelardi, Fabrizio Roccaforte, Filippo Giannazzo\**

Dr. E. Schilirò, Dr. R. Lo Nigro, S. E. Panasci, Prof. S. Agnello, Dr. F. Roccaforte, Dr. F. Giannazzo
CNR-IMM,
Strada VIII, 5 95121, Catania, Italy
e-mail: filippo.giannazzo@imm.cnr.it
        raffaella.lonigro@imm.cnr.it

Prof. S. Agnello, Prof. M. Cannas, Prof. F. M. Gelardi
University of Palermo, Department of Physics and Chemistry Emilio Segrè,
Via Archirafi 36, 90123 Palermo, Italy

Prof. S. Agnello
ATeN Center, University of Palermo,
Viale delle Scienze Ed. 18, 90128 Palermo, Italy

S. E. Panasci
Department of Physics and Astronomy, University of Catania,
Via Santa Sofia 64, 95123 Catania, Italy

**Keywords**: MoS₂, atomic layer deposition, AFM, C-AFM, Raman, Photoluminescence

**Abstract**

In this paper we demonstrated the thermal Atomic Layer Deposition (ALD) growth at 250 °C of highly homogeneous and ultra-thin (~3.6 nm) $Al_2O_3$ films with excellent insulating properties directly onto a monolayer (1L) MoS₂ membrane exfoliated on gold. Differently than in the case of 1L MoS₂ supported by a common insulating substrate ($Al_2O_3$/Si), a better nucleation process of the high-*k* film was observed on the 1L MoS₂/Au system since the ALD early stages. Atomic force microscopy analyses showed a ~50% $Al_2O_3$ surface coverage just after 10 ALD cycles, its increasing up to >90% (after 40 cycles), and an uniform ~3.6 nm film, after 80 cycles. The coverage percentage was found to be significantly reduced in the case of 2L MoS₂/Au, indicating a crucial role of the interfacial interaction between the aluminum precursor and MoS₂/Au surface. Finally, Raman spectroscopy and PL analyses provided an insight about the role played by the tensile strain and p-type doping of 1L MoS$_2$ induced by the gold substrate on the enhanced high-*k* nucleation of Al$_2$O$_3$ thin films. The presently shown high quality ALD growth of high-*k* Al$_2$O$_3$ dielectrics on large area 1L MoS$_2$ induced by the Au underlayer can be considered of wide interest for potential device applications based on this material system.

**Introduction**

Semiconducting transition metal dichalcogenides (TMDs), including MoS$_2$, WS$_2$, MoSe$_2$ and WSe$_2$, are currently widely investigated for next generation electronic and optoelectronic applications [1,2]. The deposition of high-*k* dielectrics thin films (such as Al$_2$O$_3$ and HfO$_2$) on the TMDs surface represents a key requirement for the fabrication of electronic devices [3,4,5]. As an example, the deposition of a HfO$_2$ gate insulator (~30 nm thick) on top of monolayer (1L) MoS$_2$ represented the enabling step to demonstrate a field effect transistor (FET) with excellent on/off ratio (~ 10$^8$), nearly ideal sub-threshold swing (~ 70 mV/dec) and high room temperature electron mobility (>200 cm$^2$V$^{-1}$s$^{-1}$), due to the efficient reduction of charged impurities scattering because of the high-*k* dielectric film [6]. Similarly, an Al$_2$O$_3$ top gate dielectric (~16 nm thick) was employed for the demonstration of a high mobility (~125 cm$^2$V$^{-1}$s$^{-1}$) multilayer MoS$_2$ transistor [7]. In all these cases, the high-*k* dielectric films were grown by the atomic layer deposition (ALD) technique [8], commonly employed in microelectronics to obtain uniform and conformal insulating films with a sub-nanometric control of the thickness. Ideally, layer-by-layer deposition of ultra-thin films requires the presence of a sufficiently high and uniform density of dangling bonds, necessary for precursor chemisorption on the sample surface in the early stages of the ALD process [9,10]. However, the inherent lack of out-of-plane bonds in two-dimensional (2D) layered materials represents an obstacle for an ideal ALD growth, resulting in an inhomogeneous coverage especially for very thin (<10 nm) deposited films [11,12]. In this context, relatively thick high-*k* films were employed in the first pioneering studies on $MoS_2$ transistors to achieve a uniform coverage of $MoS_2$ surface, thus minimizing the gate leakage current. However, the real application of $MoS_2$ FETs in next generation logic devices requires an aggressive scaling of the channel length and, consequently, of the high-k dielectric thickness [3]. Hence, several strategies have been investigated in the last few years to improve the ALD growth on TMDs, by tailoring the process conditions and/or by appropriate pre-functionalization treatments of the surface. Many of these approaches were inspired by ALD growth experiments performed on graphene [13,14,15]. As an example, *Park* et al. [10] systematically investigated thermal ALD of $Al_2O_3$, using trimethyl-aluminum (TMA) as the Al precursor and water ($H_2O$) as co-reactant, on the surface of different TMDs, i.e. $MoS_2$, $WS_2$ and $WSe_2$ multilayers. The deposition temperature ($T_{dep}$) and the TMA adsorption energy ($E_{ads}$) on the TMDs surface were demonstrated to play a crucial role on the uniformity of the deposited $Al_2O_3$ (~10 nm) films. In particular, $E_{ads}$, which is related to the substrate polarizability, was found to be larger for W- and Se-based TMDs (e.g., $WS_2$ and $WSe_2$) than for $MoS_2$. Furthermore, while inhomogeneous $Al_2O_3$ films (with a large density of pinholes) were obtained at typical deposition temperatures from 200 °C to 250 °C, the coverage uniformity was highly improved by lowering $T_{dep}$ to 150°C, i.e. reducing the desorption of the metal precursors from the $MoS_2$ surface [10]. Similarly to the strategy used on graphene [16], a two-steps ALD process, consisting in the low temperature (80 °C) deposition of an ultrathin AlOx layer, followed by a second ALD step at higher temperature (180 °C), was employed to obtain a homogeneous $Al_2O_3$ film with <10 nm total thickness on $MoS_2$ [17]. Despite the use of a reduced temperature at the beginning or during the whole ALD process can be beneficial to improve the coverage uniformity, it may result in a lower dielectric quality due to a reduced reactivity of the ALD-precursors [18]. Another strategy to improve the ALD growth on $MoS_2$ surface has been to replace $H_2O$ with a more reactive co-reactant, such as ozone ($O_3$), which allowed to obtain uniform $Al_2O_3$ layers (~5 nm thick) at a temperature of 200 °C [19]. Alternatively, an $O_2$-plasma pre-treatment of multilayer $MoS_2$ surface, resulting in the formation of an ultrathin Mo-oxide layer, was shown to significantly improve the uniformity of the deposited $Al_2O_3$ or $HfO_2$ films as compared to the case of pristine $MoS_2$ [20]. More recently, water plasma pre-treatments of the $MoS_2$ surface have been used to create hydroxyl groups for conventional thermal ALD at $200 - 250$ °C, resulting in the deposition of uniform $Al_2O_3$ films with thickness down to 1.5 nm [21]. In spite of these beneficial effects, the damage and chemical modifications introduced by these plasma pre-treatments can affect the electronic transport in $MoS_2$ devices. Besides thermal ALD, plasma-enhanced ALD (PEALD) processes have been also recently investigated to grow very thin films (<5 nm) of $Al_2O_3$ and $HfO_2$ on $MoS_2$ samples with different layer numbers [22,23]. In particular, electrical characterization of 1L, 2L, and 3L $MoS_2$ back-gated transistors before and after $HfO_2$ PEALD revealed the occurrence of plasma damage, resulting in significant degradation of the electronic properties especially for 1L $MoS_2$ [23].

In addition to these processes involving a chemical modification of $MoS_2$ surface, non-covalent functionalization with thin organic (e.g., perylene derivatives) [24] or inorganic (e.g., $SiO_2$ nanoparticles) [25] seeding layers has been also explored to promote the thermal ALD growth of thin $Al_2O_3$ films on $MoS_2$. However, the use of these interlayers ultimately limits the minimum thickness of the dielectric and may affect the electrical quality of the interface.

This short overview about ALD of high-$k$ dielectrics on TMDs indicates that the seeding layers and pre-functionalization approaches explored so far presents some disadvantages, while direct thermal ALD of ultra-thin films would be highly desirable. In this respect, the interaction of atomically thin $MoS_2$ layers with the underlying substrate is expected to play an important role in the ALD nucleation stage, similarly to what observed for monolayer graphene residing on some specific substrates [26,27]. As an example, *Dlubak et al.* [26] reported an enhanced $Al_2O_3$ nucleation on CVD-grown 1L graphene residing on the native metal substrates (Cu, Ni), that was ascribed to an improved ALD-precursor adsorption due the electrostatic effect of polar traps located at graphene/metal interface [28,29]. More recently, the uniform growth of ultra-thin (~2.4 nm) $Al_2O_3$ films by direct thermal ALD (at 250 °C) on monolayer epitaxial graphene on 4H-SiC(0001) has been ascribed to the beneficial effect of the carbon buffer layer at the interface with the substrate [30]. To the best of our knowledge, analogous substrate effects on the ALD nucleation onto 1L TMDs have not been reported so far.

In this paper, we investigated the ALD growth of ultra-thin (<4 nm) $Al_2O_3$ films on 1L $MoS_2$ produced by gold assisted mechanical exfoliation from bulk crystals [31,32,33,34]. This method exploits the strong Au-S interaction to exfoliate large area ($cm^2$) $MoS_2$ membranes, predominantly formed by monolayers, on a gold substrate. These high crystalline quality membranes can be subsequently transferred on insulating substrates [33,34].

Using identical ALD conditions on 1L $MoS_2$ membranes supported by gold ($MoS_2$/Au) or by $Al_2O_3$ (100 nm)/Si substrate ($MoS_2$/$Al_2O_3$/Si), the typical inhomogeneous coverage by $Al_2O_3$ islands was observed in the case of the 1L $MoS_2$/$Al_2O_3$/Si system, whereas the formation of a highly uniform $Al_2O_3$ insulating film (~3.6 nm thick) was observed on the 1L $MoS_2$/Au sample. This excellent uniformity is the result of an enhanced ALD nucleation on $MoS_2$ surface due to the interaction with Au substrate, giving rise to ~50% $Al_2O_3$ surface coverage after only 10 ALD cycles, and >90% coverage after 40 cycles. In this respect, micro-Raman and micro-photoluminescence spectroscopy analyses of 1L $MoS_2$/Au and 1L $MoS_2$/$Al_2O_3$/Si samples before and after ALD processes provided an insight on the role of the substrate-related doping and strain in the $Al_2O_3$ nucleation and growth.

**Experimental**

The gold substrate used for $MoS_2$ mechanical exfoliation was prepared by sequentially depositing a 10 nm Ni adhesion layer and a 15 nm Au film with DC magnetron sputtering on top of a $SiO_2$/Si sample. This process resulted in a very smooth Au surface with RMS roughness (<0.2 nm), suitable for the $MoS_2$ exfoliation procedure [34]. This latter was performed by pressing a bulk molybdenite stamp on the surface of a freshly prepared Au/Ni/$SiO_2$ sample, in order to avoid the adsorption of contaminants (e.g. adventitious carbon) on Au surface, which would reduce the 1L $MoS_2$ exfoliation yield [35]. This exfoliaton results in a large area mostly composed 1L MoS2 with some 2L regions identified by optical contrast, AFM and Raman spectroscopy.

The $Al_2O_3$/Si substrate used for transferring the Au-exfoliated $MoS_2$ was prepared by DC-pulsed RF reactive sputtering of 100 nm $Al_2O_3$ on a Si wafer. The transfer procedure of the large areas $MoS_2$ membranes from gold to this insulating substrate is discussed in details in Ref. [34].

Thermal ALD of $Al_2O_3$ thin films on $MoS_2$ was carried out in a PE-ALD LL SENTECH Instruments GmbH reactor, using TMA and $H_2O$ as the aluminum precursor and co-reactant, respectively. All depositions were carried out at a temperature of 250 °C and with a pressure of 10 Pa. Initially, a process consisting of 80 ALD cycles was simultaneously carried out on both 1L $MoS_2$/Au and 1L $MoS_2$/$Al_2O_3$/Si systems, to compare the $Al_2O_3$ coverage uniformity. After observing the beneficial effect of the Au substrate on the uniformity of the $Al_2O_3$ growth on 1L $MoS_2$, the nucleation and growth mechanisms on the 1L $MoS_2$/Au were investigated in a more detail, by performing shorter ALD processes (10 and 40 deposition cycles).

The surface roughness, coverage fraction and thickness of the deposited $Al_2O_3$ on $MoS_2$ were evaluated by tapping mode Atomic Force Microscopy (AFM), morphology and phase, using a DI3100 equipment by Bruker with Nanoscope V electronics. Sharp silicon tips with a curvature radius of 5 nm were used for these measurements. Furthermore, the electrical insulating properties of the very thin $Al_2O_3$ films deposited on 1L $MoS_2$/Au were evaluated by conductive AFM (C-AFM) analyses [36] using the TUNA module and Pt-coated silicon tips.

Micro-Raman spectroscopy and micro-photoluminescence (PL) measurements of $MoS_2$ on the different substrates and before/after the ALD growth of $Al_2O_3$ were carried out using an Horiba HR-Evolution micro-Raman system with a confocal microscope (100× objective) and a laser excitation wavelength of 532 nm.

## Results and discussion

Figure 1 shows the comparison between the AFM surface morphologies of $Al_2O_3$ simultaneously deposited at 250 °C by 80 ALD cycles on the surface of the 1L $MoS_2$/$Al_2O_3$/Si sample (a) and of the 1L MoS$_2$/Au sample (b), respectively. A very inhomogeneous Al$_2$O$_3$ coverage, resulting in a root mean square roughness RMS=2.5 nm, is observed on the surface of 1L MoS$_2$ transferred on the Al$_2$O$_3$/Si substrate. Furthermore, the height distribution in Fig.1 (c) shows two components, related to bare and Al$_2$O$_3$ covered MoS$_2$ areas, due to ~70% coverage and an average Al$_2$O$_3$ islands height of ~4 nm. This scenario, schematically depicted in the inset of Fig.1(c), is consistent with the commonly reported island growth during direct thermal ALD on MoS$_2$ surface. Differently, for 1L MoS$_2$ supported by the Au substrate (Fig.1(b)), a pinhole-free Al$_2$O$_3$ layer with a very flat morphology is observed after 80 ALD cycles. The deposited film exhibits a very narrow height distribution (Fig.1(d)) and low surface roughness (RMS= 0.25 nm), only slightly higher than the one measured on bare 1L MoS$_2$ on Au (RMS= 0.2 nm) [32]. Such morphological results suggest that, under identical process conditions, an enhanced Al$_2$O$_3$ nucleation occurs on the surface of 1L MoS$_2$ in contact with gold, resulting in the formation of a continuous Al$_2$O$_3$ film, as schematically depicted in the inset of Fig.1(d).

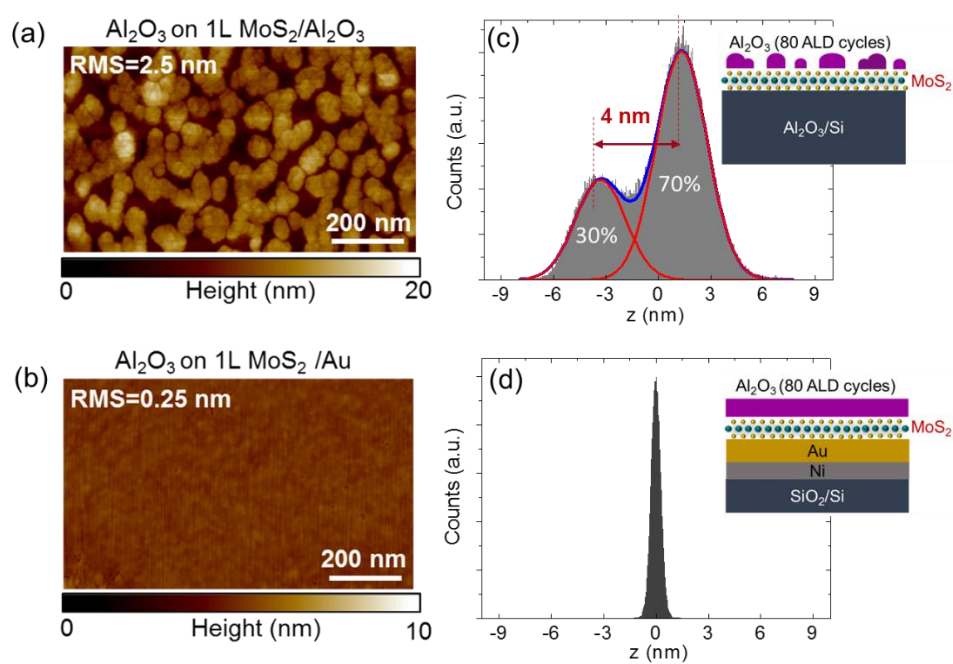

**Fig. 1**. AFM morphologies of $Al_2O_3$ simultaneously deposited at 250 °C by 80 ALD cycles on the surface of the 1L $MoS_2/Al_2O_3/Si$ sample (a) and of the 1L $MoS_2/Au$ sample (b). The root mean square (RMS) roughness values of the two samples are indicated. (c) Histogram of the height distribution obtained from the AFM map of $Al_2O_3$ on 1L $MoS_2/Al_2O_3/Si$, from which ~70% $Al_2O_3$ coverage and an average height ~4 nm of $Al_2O_3$ islands were evaluated. (d) Histogram of the height distribution for $Al_2O_3$ on 1L $MoS_2/Au$. The configuration of the deposited $Al_2O_3$ on 1L $MoS_2/Al_2O_3/Si$ and 1L $MoS_2/Au$ is schematically illustrated in the insets of panels (c) and (d).

In order to evaluate the thickness and the electrical insulating quality of the uniform $Al_2O_3$ film deposited on 1L $MoS_2/Au$, C-AFM morphology and current maps were simultaneously acquired by scanning the metal tip across a step between the $Al_2O_3/1L$ $MoS_2$ stack and the underlying Au substrate, as schematically depicted in Fig.2(a). Fig.2(b) shows a morphological image collected in the proximity of a crack in the 1L $MoS_2$ membrane. The growth of a uniform and compact $Al_2O_3$ film on 1L $MoS_2$ and a poor ALD growth on the bare Au surface can be deduced from this image. Furthermore, the height line-profile in Fig.2 (c) displays a total thickness of the $Al_2O_3/MoS_2$ stack of ~4.3 nm, from which a deposited $Al_2O_3$ thickness of ~3.6 nm can be estimated, by subtracting the thickness of 1L $MoS_2$ on Au (~0.7 nm) [34]. The slightly increased height observed at the step edge can be ascribed to the folding of the broken 1L $MoS_2$ membrane. The very good electrical insulating properties of the 3.6 nm $Al_2O_3$ film deposited onto 1L $MoS_2$ on Au are demonstrated by the current map in Fig.2(d), acquired by applying a bias of 3 V between the tip and the gold substrate. In particular, the line-profile in Fig.2(e) shows very low current values (between 10 and 30 fA) in the $Al_2O_3/1L$ $MoS_2$ region, whereas the saturation value of the current sensor was reached on Au region. The current conduction in the region close to the step edge suggests lower insulating properties of $Al_2O_3$ deposited on the locally folded $MoS_2$ membrane.

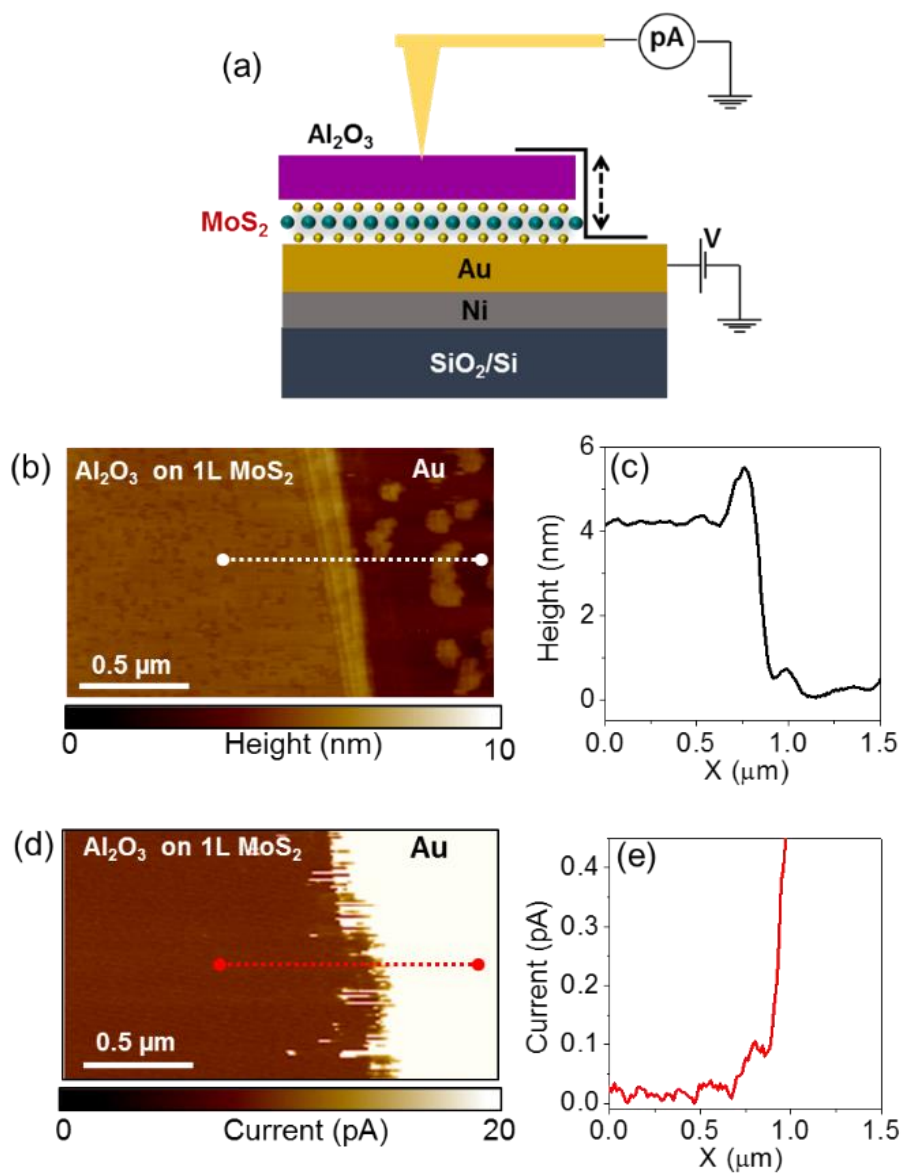

**Fig.2** (a) Schematic of the step between the $Al_2O_3/1L$ $MoS_2$ stack and the underlying Au substrate. (b) AFM image and (c) height line-profile of the step, from which a deposited $Al_2O_3$ thickness of ~3.6 nm was estimated, after subtracting 1L $MoS_2$ thickness (~0.7 nm) (d) C-AFM current map simultaneously acquired with a bias V=3V and (e) current profile, demonstrating a good insulating quality of the deposited $Al_2O_3$ film onto 1L $MoS_2$ on Au.

After demonstrating the formation of a homogeneous ~3.6 nm $Al_2O_3$ insulating film on top of 1L $MoS_2$/Au by 80 ALD cycles at 250 °C, we investigated the film nucleation and growth stages by AFM analyses performed after a reduced number of ALD cycles at the same temperature.

Fig.3(a) and (d) show two tapping mode morphological images acquired on 1L $MoS_2$/Au samples after 10 and 40 ALD cycles, respectively. In particular, after 10 cycles, a very irregular and ultrathin coating can be deduced from the morphological image, resulting in a RMS≈0.4 nm, slightly higher than the ~0.2 nm value measured on the bare 1L $MoS_2$/Au sample. On the other hand, a grain-shaped morphology of the deposited $Al_2O_3$ film can be clearly observed after 40 ALD cycles (see Fig.3(d)), suggesting the occurrence of 3D growth of $Al_2O_3$ islands on top of the inhomogeneous nucleation layer formed at lower number of cycles. A quantification of the coverage percentage is very difficult from these morphological images. On the other hand, the $Al_2O_3$ coated and uncoated 1L $MoS_2$ areas can be clearly distinguished in the corresponding AFM phase maps (Fig.3(b) and (e)), as the phase signal is known to be very sensitive to the surface properties of materials. In particular, the red and black contrast in these two images correspond to the $Al_2O_3$-covered and uncovered 1L $MoS_2$, respectively. Furthermore, the histograms of the phase distribution extracted from the two maps are reported in Fig.3(c) and (f), from which an $Al_2O_3$ coverage percentage of 50% and 93% were evaluated after 10 and 40 ALD cycles, respectively. This very high coverage after only 40 ALD cycles, corresponding to ~1.4 nm $Al_2O_3$ thickness (measured by an AFM step-height analysis as in Fig.2(b)), demonstrates a very good nucleation on Au supported 1L $MoS_2$.

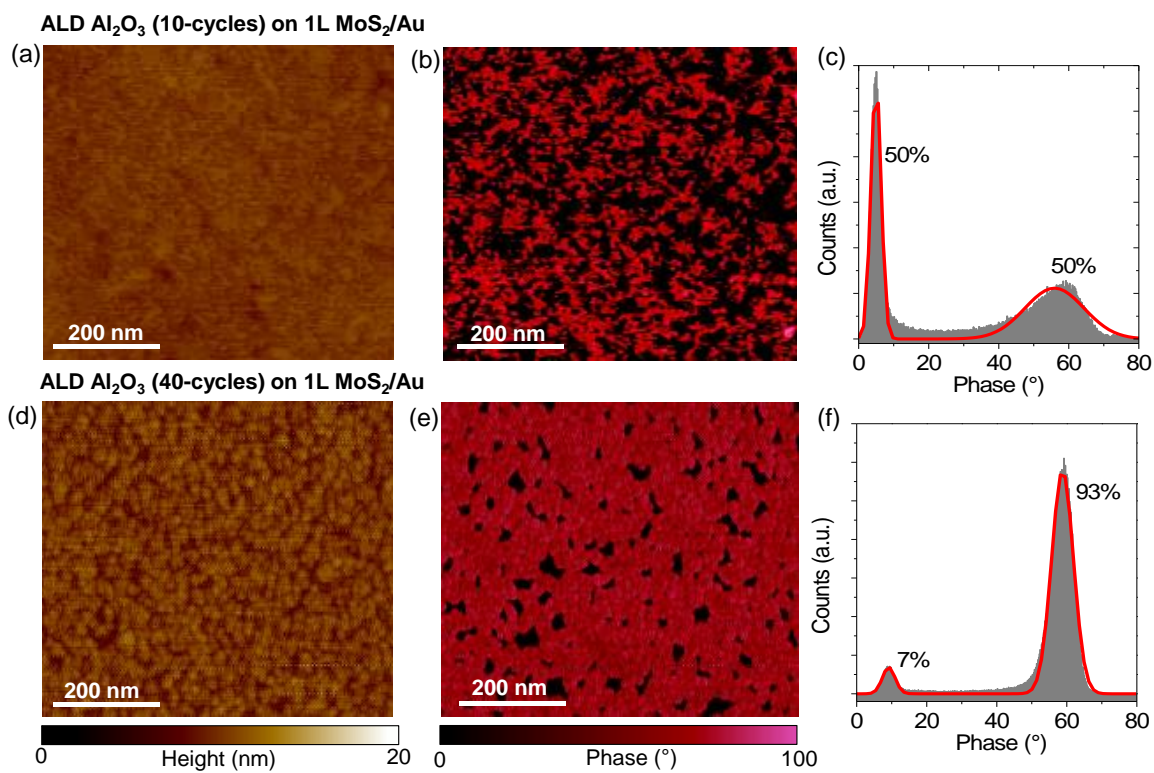

**Figure 3** (a) AFM morphology and (b) phase map measured after 10 $Al_2O_3$ ALD cycles on 1L $MoS_2$/Au. (c) Histogram of the phase distribution and evaluation of the $Al_2O_3$ coverage percentage on the same sample. (d) AFM morphology and (e) phase map measured after 40 $Al_2O_3$ ALD cycles on 1L $MoS_2$/Au. (f) Histogram of the phase distribution and evaluation of the $Al_2O_3$ coverage percentage on the same sample

Interestingly, the coverage degree was found to be strongly dependent on the number of $MoS_2$ layers on which the ALD process is performed. Fig.4 shows a representative AFM morphology (a) and phase image (b) collected after 40 ALD cycles in a region of the exfoliated $MoS_2$ membrane on Au including 1L and 2L $MoS_2$ areas. A very different $Al_2O_3$ coverage can be clearly observed (especially in the phase image) on 1L and 2L $MoS_2$ regions, with a much denser $Al_2O_3$ nucleation on 1L than on the 2L $MoS_2$.

This observation further confirms the key role played by the interaction at $MoS_2$/Au interface on the ALD nucleation on top of $MoS_2$. As schematically illustrated in Fig.4(c), left image, the strongly polarized Au-S bond is expected to significantly affect the electronic and vibrational properties of the first $MoS_2$ layer closely in contact with the substrate, which can be responsible of the observed enhanced ALD nucleation on 1L-MoS$_2$. On the other hand, in the case of 2L MoS$_2$ (Fig.4(c), right image), the Au-S interaction is expected to be partially screened by the presence of the first MoS$_2$ layer, resulting in a less-efficient ALD growth. A similar degradation of Al$_2$O$_3$ coverage homogeneity has been also observed in the case of bilayer or few-layers CVD graphene on the native copper substrate [26] and of bilayer epitaxial graphene on 4H-SiC(0001) [27].

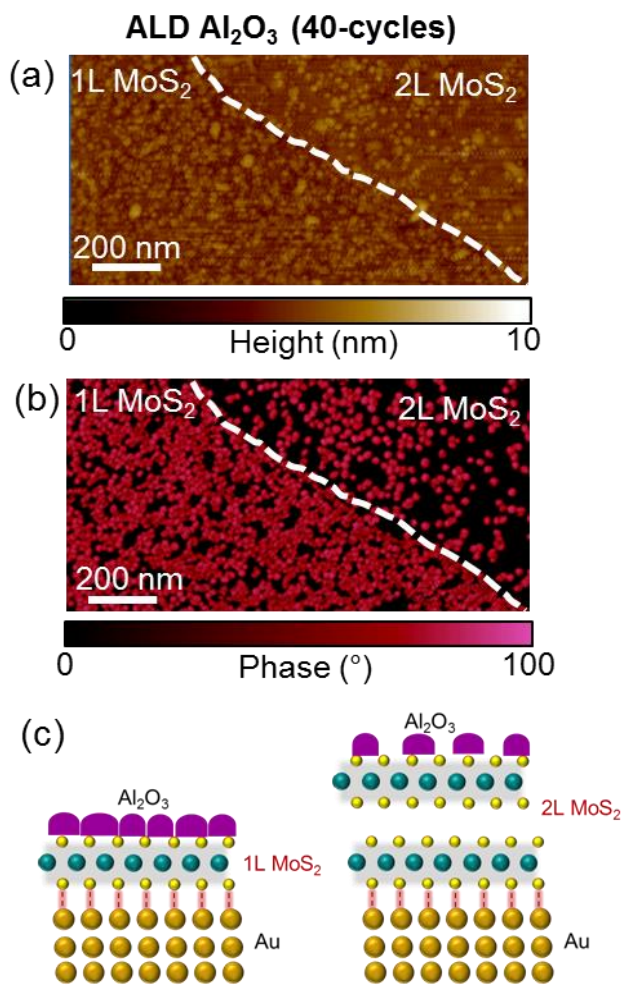

**Fig.4**: AFM-morphology (a) and phase image (b) of Al$_2$O$_3$ deposited by 40 ALD cycles on a region including both 1L and 2L MoS$_2$ on Au. (c) Schematic illustration of the impact of Au-S interaction on the ALD nucleation on 1L- and 2L MoS$_2$.

Then, micro-Raman spectroscopy analyses have been performed to investigate the strain and doping status of 1L MoS$_2$ residing on Au and Al$_2$O$_3$ substrates before and after the ALD growth. Fig.5(a) reports two representative Raman spectra for as-exfoliated 1L MoS$_2$ on Au (reference) and after 80

TMA/$H_2O$ ALD cycles, resulting in the homogeneous ~3.6 nm $Al_2O_3$ film deposition shown in Fig. 2(b). The corresponding Raman spectra for 1L $MoS_2$ transferred onto the $Al_2O_3$/Si substrate (reference) and after the 80 ALD cycles are shown in Fig.5(b). From the comparison of the reference spectra on the two substrates, a significantly higher separation $\Delta\omega$ between the in-plane (E') and out-of-plane ($A_1$') vibrational peaks is observed for 1L $MoS_2$ on Au ($\Delta\omega \approx 21$ cm$^{-1}$) as compared to the case of 1L $MoS_2$ on $Al_2O_3$/Si ($\Delta\omega \approx 18$ cm$^{-1}$). Such a difference is due to significant red-shift of the E' peak (mostly associated to the strain) and to a slight blue shift of the $A_1$' peak (mostly related to the doping) for 1L $MoS_2$ membrane on gold. Interestingly, the E' peak red-shift is further increased and the $A_1$' peak blue-shift is slightly reduced after ALD of the uniform $Al_2O_3$ film on the gold supported membrane. On the other hand, only a slight red shift of the $A_1$' peak was observed after 80 ALD cycles on the $Al_2O_3$ supported 1L $MoS_2$, probably due to the inhomogeneous $Al_2O_3$ coverage (as shown in Fig.1(a)). In order to achieve a quantification of the strain $\varepsilon$ (%) and doping n (cm$^{-2}$) for the 1L $MoS_2$ membranes on the two different substrates before and after the ALD process, a correlative analysis of the $A_1$' vs E' peak frequencies has been carried out in Fig.5(c), according to the procedure recently discussed in Ref. [34]. The red and black lines in Fig.5(c) represent the theoretical behaviour of the peaks' frequencies for 1L $MoS_2$ subjected only to a biaxial strain (tensile or compressive) or to doping (n-type or p-type), respectively. The crossing point (gray square) of these lines corresponds to literature values of the E' and $A_1$' positions for a free-standing 1L $MoS_2$, taken as the best approximation for ideally unstrained and undoped $MoS_2$. The spacing of the dashed lines parallel to the ideal strain and doping lines is associated with carrier density changes of $0.1 \times 10^{13}$ cm$^{-2}$ and strain changes of 0.1%, respectively. The black point is the average value of the $A_1$' vs E' frequencies from several (>20) Raman spectra measured on the reference 1L $MoS_2$/Au sample, whereas the error bars are the standard deviations from this statistical analysis. The magenta point is the average value obtained from Raman analyses on several points of the reference 1L $MoS_2$/$Al_2O_3$ sample. For this sample, the $A_1$' frequency exhibits a significant dispersion (indicated by the error bar), whereas the small E' frequency dispersion is within the data point. According to the graphical analysis in Fig.5(c), the reference 1L $MoS_2$/Au sample is characterized by an average tensile strain of $\varepsilon \approx 0.21\%$ and p-type doping of $n \approx -0.25 \times 10^{13}$ $cm^{-2}$, whereas an opposite compressive strain $\varepsilon \approx -0.25\%$ and n-type doping $n \approx 0.5 \times 10^{13}$ $cm^{-2}$ are observed for 1L $MoS_2$ transferred onto the $Al_2O_3$/Si substrate. Such n-type behaviour is consistent with the unintentional doping type commonly reported for exfoliated or CVD-grown $MoS_2$, which has been associated to the presence of defects (e.g. sulphur vacancies) or to other impurities in the $MoS_2$ lattice [37,38]. In the case of 1L $MoS_2$ on Au, a strong electron transfer to the substrate is guessed, which overcompensates the native n-type doping, resulting in a net p-type behaviour. Furthermore, the tensile strain for 1L $MoS_2$ on Au can be ascribed to the lattice mismatch between $MoS_2$ and the Au surface, mostly exposing (111) orientation [39,40]. The opposite strain and doping status of 1L $MoS_2$/$Al_2O_3$ and 1L $MoS_2$/Au can be the explanation of the very different $Al_2O_3$ coverage observed in the two systems under identical thermal ALD conditions.

Noteworthy, the blue point in Fig.5(c) obtained from Raman analyses after 80 ALD cycles on 1L $MoS_2$/Au indicates a further increase of the tensile strain ($\varepsilon \approx 0.65\%$), as compared to the original value of 0.21%, which can be associated to the formation of a compact $Al_2O_3$ film on top of $MoS_2$. On the other hand, no significant changes with respect to the original p-type doping value was observed after the ALD growth, confirming that the doping status of the film is strongly dominated by the strong Au-1L $MoS_2$ interaction. Finally, the data-point for the 1L $MoS_2$/$Al_2O_3$ sample after the 80 ALD cycles indicates no significant changes in the tensile strain, consistently with the highly inhomogeneous $Al_2O_3$ coverage, and an increase of the n-type doping to $n \approx 0.6 \times 10^{13}$ $cm^{-2}$. This latter can be ascribed to positively charged defects [41] at the interface between the poor quality $Al_2O_3$ film and 1L $MoS_2$.

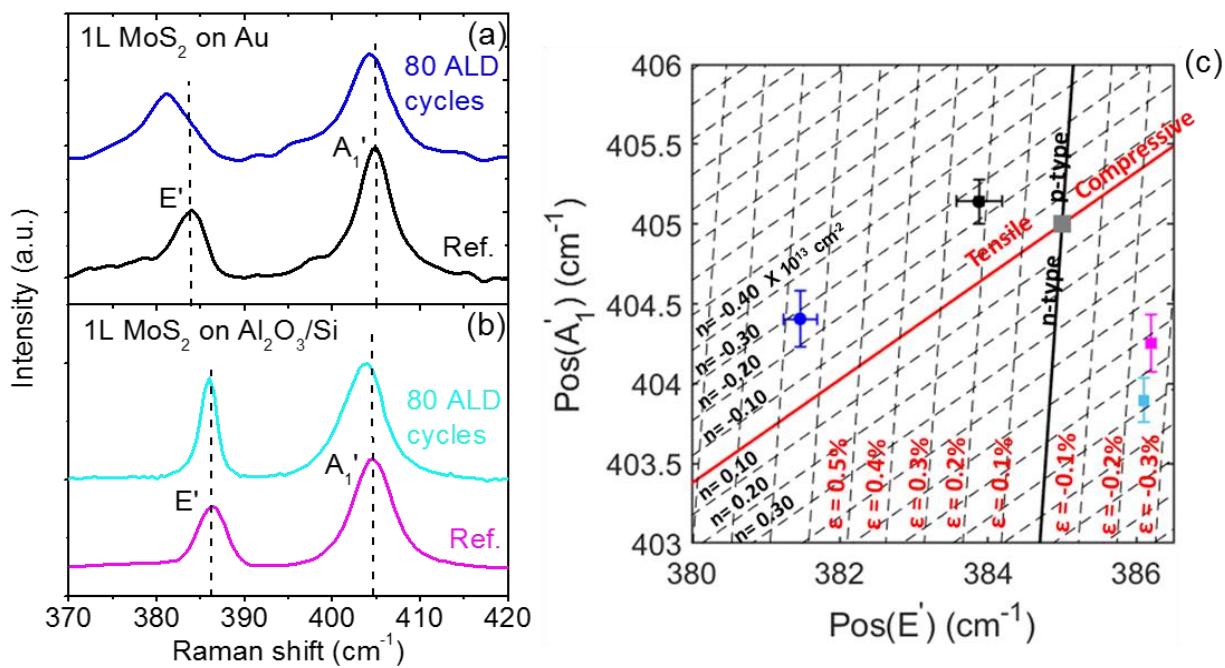

**Figure 5** Representative Raman spectra collected on 1L $MoS_2$ on Au (a) and 1L $MoS_2$ on $Al_2O_3$/Si (b) before (reference) and after 80 $Al_2O_3$ ALD cycles. (c) Correlative plot of the $A_1$' vs E' peak frequencies of the Raman spectra acquired on the 1L $MoS_2$ on Au before (black circle) and after (blue circle) ALD deposition and for 1L $MoS_2$ on $Al_2O_3$/Si before (magenta square) and after (cyan square) ALD deposition , allowing to evaluate the type and average values of strain and doping of 1L $MoS_2$ membrane.

Raman analyses clearly show very different strain and doping properties of 1L MoS$_2$ residing on the Au and Al$_2$O$_3$/Si substrates, which finally result in the growth of different quality Al$_2$O$_3$ films after ALD deposition.

In Fig.6, micro-PL spectra acquired on the two reference 1L MoS$_2$ samples and after 80 cycles ALD growth are also reported, to further elucidate the impact of the substrate and of the deposition process on the optical emission properties of the direct bandgap 1L MoS$_2$ membrane.

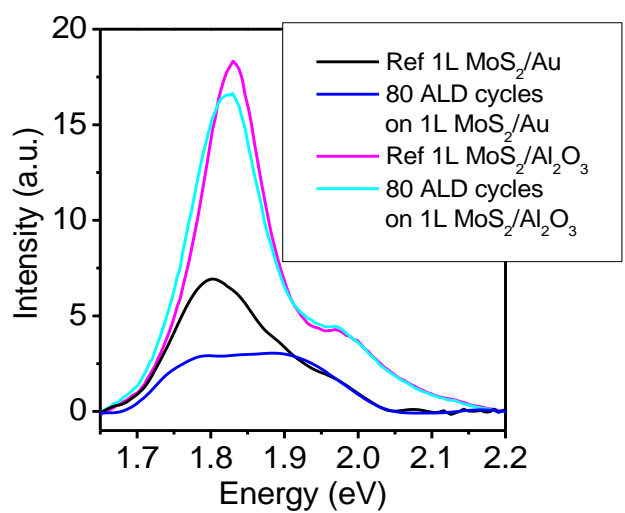

**Figure 6** Micro-PL spectra acquired on 1L MoS$_2$/Al$_2$O$_3$ and on 1L MoS$_2$/Au samples before (ref. spectra) and after 80 Al$_2$O$_3$ ALD cycles.

A prominent emission peak located at 1.84 eV can be observed for 1L MoS$_2$ supported by Al$_2$O$_3$/Si, whereas a significant reduction of the PL intensity accompanied by the red shift of the main peak position at 1.79 eV is found for the reference sample on Au. A similar quenching of the PL intensity has been reported for 1L MoS$_2$ exfoliated on Au [34,35] and for MoS$_2$ functionalized with Au nanoparticles [42]. This behavior can be explained in terms of a preferential transfer of photoexcited charges from MoS$_2$ to Au. In addition, the tensile strain of 1L MoS$_2$ in contact with Au can also play a role in the reduction of the PL yield [43]. After 80 ALD cycles, only a small reduction of the PL intensity was observed for the 1L MoS$_2$/Al$_2$O$_3$ sample, which can be explained by the highly inhomogeneous Al$_2$O$_3$ coverage and to the small interaction of MoS$_2$ with the dielectric substrate. On the other hand, further quenching of the PL intensity was found in the case of 1L MoS$_2$/Au sample covered by the ~3.6 nm uniform Al$_2$O$_3$ film. This observation can be consistent with the increase of the tensile strain observed by Raman analyses and to the increase of the interaction with the Au substrate due to the material added on top of MoS$_2$.

**Conclusion**

In conclusion, we have demonstrated the direct thermal ALD growth at 250 °C of highly homogeneous and ultra-thin (~3.6 nm) Al$_2$O$_3$ films with excellent insulating properties onto a monolayer MoS$_2$ membrane exfoliated on gold. Differently from the case of 1L MoS$_2$ supported by a common insulating substrate (Al$_2$O$_3$/Si), an enhanced nucleation of the high-*k* films was observed on the 1L MoS$_2$/Au system since the early stages of the ALD process, with the Al$_2$O$_3$ surface coverage increasing from ~50% (after only 10 ALD cycles) to >90% (after 40 cycles), up to the final ~3.6 nm uniform film (after 80 cycles). The coverage percentage was found to be significantly reduced in the case of 2L MoS$_2$/Au, indicating a crucial role of the S-Au interaction at the interface in the observed phenomena. Raman spectroscopy and PL analyses provided an insight about the role played by the tensile strain and p-type doping of 1L MoS$_2$ induced by the gold substrate on the enhanced high-*k* nucleation on MoS$_2$ surface.

The demonstrated high quality ALD growth of high-*k* dielectrics on large area 1L MoS$_2$ produced by the Au-assisted exfoliation can find important device applications, including the fabrication of FETs by optimized transfer of the Al$_2$O$_3$/1L MoS$_2$ stack onto an insulating substrate, or the passivation of non-volatile switching memory devices based on Au/1L MoS$_2$/Au junctions.

**Acknowledgements**

The authors acknowledge S. Di Franco (CNR-IMM) for the expert assistance in the sample preparation, P. Fiorenza and G. Greco (CNR-IMM) for useful discussions. This paper has been supported, in part, by MUR in the framework of the FlagERA- 732JTC 2019 project "ETMOS". E.S.

WILEY-VCH

acknowledges the PON project EleGaNTe (ARS01_01007) for financial support. Part of the experiments have been carried out using the facilities of the Italian Infrastructure Beyond Nano.